\documentclass[aps,prl,preprint,groupedaddress]{revtex4}
\usepackage{graphicx}
\usepackage{amssymb}
\usepackage{epstopdf}
\begin{document}

\title{Theory of Inelastic Electron Tunneling from a Localized Spin in the Impulsive Approximation}
\author{M. Persson}
\affiliation{The Surface Science Research Centre, The University of Liverpool, Liverpool, L69 3BX, UK}

\begin{abstract}
A simple expression for the conductance steps in the inelastic electron tunneling from spin excitations in a single  magnetic atom adsorbed on a non-magnetic metal surfaces is derived. The inelastic coupling between the tunneling electron and the spin is via the exchange coupling and is treated in an impulsive approximation using the Tersoff-Hamann approximation for the tunneling between the tip and the sample. Our results for conductance steps justify the analysis carried out by Hirjebedin et al. [Science {\bf 317}, 1199 (2007)]  of observed step-like conductances by inelastic electron tunneling from spin excitations in a single magnetic adatom using a simple spin matrix element. In addition, our result gives a simple expression for the magnitudes of conductance steps and their lateral spatial variation with respect to the tip position, which can be calculated directly from spin-polarized wave functions at the Fermi level of the sample. 
\end{abstract}

\maketitle
A recent breakthrough in the field of atomic-scale magnetism of surfaces was made by the observation of inelastic electron tunneling (IET) of spin-excitations in single and chains of magnetic atoms adsorbed on a surface using a low temperature scanning tunneling microscope (STM) in the presence of a high magnetic field~\cite{Heietal04,Hiretal06,Hiretal07}.  The spin-excitations show up as steps in the measured tunneling conductance with respect to the voltage across the tip and sample. An analysis of the observed conductance steps in terms of simple spin-model Hamiltonians has revealed detailed atomic-scale information about the magnetic properties of adsorbed magnetic atoms on surfaces and their magnetic interactions~\cite{Hiretal06,Hiretal07}. This analysis showed also the presence of selection rules in IET.  Furthermore, in the case of single magnetic adatoms the relative magnitude of the conductance steps could be accounted for by a simple spin matrix element taken from inelastic neutron scattering~\cite{Hiretal07}. The detailed physical mechanisms behind IET from spin-excitations and what determines the magnitudes of the observed conductance steps are still unclear and need to be resolved.

In this note we develop a simple single-electron theory for IET from localized spin excitations of a single magnetic adsorbate on a non-magnetic substrate based on the Tersoff-Hamann approximation for elastic tunneling~\cite{TerHam83,TerHam85}. The spin of the adsorbate is treated in the adiabatic approximation. The inelastic coupling between the tunneling electron and the spin  is via the exchange interaction. This coupling is treated in a non-perturbative, impulsive approximation that was first developed in inelastic scattering in nuclear physics~\cite{CheWic52} and later applied to rotational scattering in molecular physics~\cite{AbrHer69}. We derive a simple expression for the conductance steps whose magnitudes are determined by the spin matrix elements used by Hirjibehedin~\cite{Hiretal07} and the local density of states of spin-averaged and -polarized parts of the one-electron functions.

In developing our theory for IET from spin excitations, we consider specifically a localized, collinear spin moment of a single adsorbed magnetic atom on a non-magnetic metal substrate. The electronic structure is modeled in an one-electron approximation as provided by density functional theory. We begin by characterizing the electronic states of the sample as a function of the spin direction with boundary conditions specified along a fixed quantization direction. Next we introduce the Tersoff-Hamann approximation for the differential tunneling conductance between the sample with a fixed spin direction and a non-magnetic tip. Before generalizing the TH approximation to IET from the localized spin excitation we present the adiabatic approximation for the spin dynamics. We end by a short discussion of the results.

The localized spin moment of the magnetic adatom arises from an exchange interaction between the electrons. In discussing the form of the electronic states, we neglect the small interaction terms arising from any external magnetic field and spin-orbit interaction. The one-electron states of magnetic adatom with a spin moment along a direction $\hat{n}$ is then obtained by a spin rotation of the states when the spin is along the fixed spin quantization direction $\hat{z}$. A spin moment distribution $s(\vec{r})\hat{z}$ oriented along $\hat{z}$ is obtained in an one-electron approximation by a spin-dependent one-electron potential,
\begin{eqnarray}
\bar{v}(\vec{r}) & = & v_{00}(\vec{r})\bar{I} + \Delta v_{0}(\vec{r})\bar{I} 
                                  + \Delta v_s(\vec{r})\sigma_z \\
                        & = & v_0(\vec{r})\bar{I} + \Delta v_s(\vec{r})\sigma_z 
\label{eq:vdef}
\end{eqnarray}
where $\sigma_z$ is a Pauli spin matrix. Here $v_{00}(\vec{r})\bar{I}$ is the spin-independent potential of the bare substrate and $\Delta v_{0}(\vec{r})\bar{I}  + \Delta v_s(\vec{r})\sigma_z$ is the localized adsorbate-induced potential. Far away from the adsorbate in the bulk region, the one-electron potential is spin-independent, $v(\vec{r}) \simeq v_{00}(\vec{r})\bar{I}$ resulting in two-fold degenerate scattering states $|\psi_{\mu,\sigma^\prime}\rangle$ with energy $\epsilon_\mu$. These states are determined by the Lippmann-Schwinger equation using the adsorbate-induced potential, the one-electron Green function of the bare non-magnetic surface and incident spin-degenerate states in the bulk region of the substrate, 
\begin{equation}
\langle \vec{r}, \sigma|\phi_{\mu,\sigma^\prime}\rangle =
 \delta_{\sigma,\sigma^\prime}\langle \vec{r}|\psi_{\mu}^{\rm in}\rangle \ ,
\label{eq:StatBoun}
\end{equation}
The states are labeled by an index $\mu$ and the spin index $\sigma^\prime=\pm1$ and they are collinear
\begin{equation}
\langle \vec{r}, \sigma|\psi_{\mu,\sigma^\prime}\rangle = 
 \delta_{\sigma,\sigma^\prime}\psi_{\mu,\sigma^\prime}( \vec{r}) .
 \end{equation}
To ease the notation when discussing the form of the electronic states for an arbitrary direction of the spin moment, we pair the two degenerate spinor wave functions $\langle \vec{r}, \sigma|\psi_{\mu,\sigma^\prime}\rangle$ into 2x2 matrices as,
\begin{equation}
\bar{\psi}_{\mu;\sigma\sigma^\prime} ( \vec{r})= \langle \vec{r},\sigma |\psi_{\mu,\sigma^\prime}\rangle
\label{eq:StateNot}
\end{equation}
and these collinearw ave functions can now be represented in the following compact form
\begin{eqnarray}
\bar{\psi}_{\mu} ( \vec{r}) & = & \bar{\psi}^{\rm in}_{\mu} ( \vec{r}) + \bar{\psi}^{\rm sc}_{\mu} ( \vec{r}) \label{eq:StaFora} \\
 & = & \psi_{0\mu} ( \vec{r}) \bar{I}+\psi_{s\mu} ( \vec{r})\sigma_z \ .
\label{eq:StaFor}
\end{eqnarray}
Here the incident wave function $\bar{\psi}^{\rm in}_{\mu}(\vec{r})$ in Eq. (\ref{eq:StaFora}) is the compact form of the wave function in Eq.(\ref{eq:StatBoun}),
 \begin{equation}
\bar{\psi}^{\rm in}_{\mu} ( \vec{r}) =\psi_{\mu}^{\rm in} ( \vec{r}) \bar{I} \ .
\label{eq:StaIn}
\end{equation}
and $\bar{\psi}^{\rm sc}_{\mu} ( \vec{r})$ is the scattered part of the wave function from the adsorbate-induced potential $\Delta v_{0}(\vec{r})\bar{I}  + \Delta v_s(\vec{r})\sigma_z$.  In Eq.(\ref{eq:StaFor}), $\psi_{0\mu} ( \vec{r})$ and $\psi_{s\mu} ( \vec{r})$ are the spin-averaged and -dependent parts of the wave functions, defined as,
\begin{eqnarray}
\psi_{0\mu} ( \vec{r}) & = & \frac{1}{2}
(\psi_{\mu,\sigma=1}( \vec{r})+\psi_{\mu,\sigma=-1}( \vec{r})) \\
\psi_{s\mu} ( \vec{r}) & = & \frac{1}{2}
(\psi_{\mu,\sigma=1}( \vec{r})-\psi_{\mu,\sigma=-1}( \vec{r}))
\label{eq:psi0sDef}
\end{eqnarray}
In particular,  the spin moment distribution $s(\vec{r})\hat{z}$ is determined by these wave functions as,
\begin{equation}
s(\vec{r}) = 4\sum_{\mu: \epsilon_\mu < \epsilon_F}
\Re\{\psi^\star_{0\mu} ( \vec{r})\psi_{s\mu} ( \vec{r})\}
\label{eq:SMom}
\end{equation}
Note that the relative phases of the wave functions 
$\psi_{\mu} ( \vec{r})$ in Eq.~(\ref{eq:psi0sDef})  are fixed by the boundary condition in Eq.~(\ref{eq:StaIn}).

Now we consider the spin-moment to be aligned along an arbitrary direction $\hat{n}$ and the corresponding one-electron potential $\bar{v}(\vec{r};\hat{n})$ is simply obtained by a spin-rotation using the matrix $U(\hat{n})$ that rotates a spin state $|\sigma=1\rangle_s \equiv |\hat{z}\rangle_s$ to $|\hat{n}\rangle_s$: $|\hat{n}\rangle_s=U(\hat{n})|\hat{z}\rangle_s$ and 
\begin{eqnarray}
\bar{v}(\vec{r};\hat{n}) & = & U(\hat{n}) \bar{v}(\vec{r})U^\dagger(\hat{n})  \\
& = & v_0(\vec{r})I + \Delta v_s(\vec{r})\hat{n}.\vec{\sigma} ,
\label{eq:vrot}
\end{eqnarray}
where we in the last step have used $U(\hat{n})\sigma_zU^\dagger(\hat{n})=\hat{n}.\vec{\sigma}$. In the notation introduced in Eq.(\ref{eq:StateNot}), the new stationary states $|\psi_{\mu,\sigma}(\hat{n})\rangle$ in the potential $\bar{v}(\vec{r};\hat{n})$ that obeys the incoming boundary conditions set by the states in Eq. (\ref{eq:StaIn}) are then obtained by the spin rotation,
\begin{eqnarray}
\bar{\psi}_{\mu} ( \vec{r};\hat{n}) & = & 
U(\hat{n})\bar{\psi}_{\mu} ( \vec{r})U^\dagger(\hat{n})  \label{eq:BarPsiDef}\\
& = & \psi_{0\mu} ( \vec{r}) \hat{I}+\psi_{s\mu} ( \vec{r})\hat{n}.\vec{\sigma} .
\label{eq:BarPsiRes}
\end{eqnarray}

Now we are in position to obtain the tunneling conductance between a non-magnetic tip and the magnetic sample with a fixed spin direction $\hat{n}$ using the TH approximation. This approximation is based on the Bardeen approximation for the tunneling amplitude and a $s$ wave approximation for the electrons tunneling from the tip. The differential tunneling conductance $G(V)$ at small positive sample biases $V$ and at zero-temperature is then given by~\cite{TerHam85},
\begin{equation}
G(V) = \frac{2\pi e^2}{\hbar} \sum_{\mu,\sigma^\prime; \nu,\sigma}
|T_{\mu\sigma^\prime;\nu\sigma}(\hat{n})|^2\delta(\epsilon_\mu -\epsilon_F - eV)\delta(\epsilon_\nu -\epsilon_F)
\label{eq:GDef}
\end{equation}
where $\epsilon_F $ is the Fermi energy and the tunneling (transition) amplitude $T_{\mu\sigma^\prime;\nu\sigma}(\hat{n})$ from a tip state $|\psi_{\nu,\sigma}\rangle$ to a sample state $|\psi_{\nu,\sigma^\prime}(\hat{n})\rangle$ is given by
\begin{equation}
T_{\mu\sigma^\prime;\nu\sigma} (\hat{n}) = t_s  \langle\psi_{\mu\sigma^\prime}(\hat{n}) |\vec{r}_0,\sigma\rangle
\langle s|\psi_\nu\rangle \ .
\label{eq:TunAmp}
\end{equation}
Here $\langle s|\psi_\nu\rangle$ is the spin-independent $s$ partial-wave amplitude around the tip apex $\vec{r}_0$ for a tip state $|\psi_{\nu,\sigma}\rangle$ and $t_s$ is determined by the tip. Note that here should the sample scattering states satisfy outgoing boundary conditions instead of the incoming boundary conditions in Eq. (\ref{eq:StatBoun}). Inserting the tunneling amplitude in Eq. (\ref{eq:TunAmp}) into Eq. (\ref{eq:GRes}), one recovers the TH result that the differential conductance is proportional to the spin-averaged local density of states (LDOS) at the tip apex as,
\begin{equation}
G(V) = C_t {\mathrm Tr}[\bar{\rho}(\epsilon_F + eV;\vec{r}_0,\hat{n})]
\label{eq:GRes}
\end{equation}
where ${\mathrm Tr}$ is the trace over the spin components and 
\begin{equation}
\bar{\rho}_{\sigma^\prime\sigma}(\epsilon;\vec{r}_0,\hat{n}) = \sum_{\mu,\sigma^{\prime\prime}} \langle \vec{r}_0,\sigma^\prime|\psi_{\mu\sigma^{\prime\prime}},\hat{n}\rangle\langle\psi_{\mu\sigma^{\prime\prime}},\hat{n} |\vec{r}_0,\sigma\rangle\delta(\epsilon - \epsilon_\mu)
\label{eq:rhobarDef}
\end{equation}
and all tip quantities appear in $C_t$ defined as,
\begin{equation}
C_t =\frac{2\pi e^2}{\hbar} |t_s|^2\sum_\nu|\langle s|\psi_\nu\rangle|^2\delta(\epsilon_\nu-\epsilon_F) \ .
\label{eq:CtDef}
\end{equation}
The spin indices are avoided explicitly using the matrix notation introduced in Eq. (\ref{eq:StateNot}) and Eq.(\ref{eq:rhobarDef}) simplifies to
\begin{equation}
\bar{\rho}(\epsilon;\vec{r}_0,\hat{n}) = \sum_{\mu} \bar{\psi}_{\mu} ( \vec{r}_0;\hat{n}) \bar{\psi}^\dagger_{\mu} ( \vec{r}_0;\hat{n}) \delta(\epsilon - \epsilon_\mu) \ .
\end{equation}
The result for $\bar{\psi}_{\mu} ( \vec{r};\hat{n})$ in Eq.(\ref{eq:BarPsiDef}) and the simple properties of the Pauli matrices give directly the spin-averaged LDOS as,
\begin{equation}
{\mathrm Tr}[\bar{\rho}(\epsilon;\vec{r},\hat{n})] = 
2\sum_{\mu}( |\psi_{0\mu} ( \vec{r})|^2 + |\psi_{s\mu} ( \vec{r})|^2)\delta(\epsilon - \epsilon_\mu) .
\end{equation}
Thus from Eq.(\ref{eq:GRes}), one finds that differential tunneling conductance $G(V)$ is independent of the spin-direction $\hat{n}$ as expected from spin-rotation invariance. Note that the tunneling amplitudes in Eq.(\ref{eq:TunAmp}) with respect to $\hat{z}$ is dependent on $\hat{n}$ . After presenting the model of the spin-dynamics, we will show that this dependence gives rise to spin-excitations.

The localized spin of a single adsorbed magnetic atom on a non-magnetic metal substrate is treated in the adiabatic approximation. The motion of the spin moment distribution is assumed to be rigid and the electrons are assumed to follow instantaneously the motion of the spin direction $\hat{n}$. Thus the ground state energy of the electrons for a fixed spin direction $\vec{S} = \sqrt{S(S+1)}\hat{n}$ determines the effective Hamiltonian $\mathcal{H}_{eff}$ for the spin motion. A perturbative treatment of the weak spin-orbit interaction and the weak external magnetic field $\vec{B}$ gives in general rise to an $\mathcal{H}_{eff}$ that contains terms up to second order in $\vec{S}$ and $\vec{B}$. The $2S+1$ spin eigen states $|\Phi_n\rangle$ of $\mathcal{H}_{eff}$ with eigen-energies $E_n$ are here simply labeled by their increasing value with an integer $n$. For instance, in the case discussed by Hirjebedin and coworkers~\cite{Hiretal07}, this Hamiltonian has the form
\begin{equation}
\mathcal{H}_{eff} = g\mu_B\vec{B}\cdot\vec{S} + DS_z^2+ E(S_x^2-S_y^2)
\end{equation}
where $g$ is the gyromagnetic ratio, $\mu_B$ is the Bohr magneton, $D$ and $E$ are magnetic anisotropy parameters arising from the spin-orbit interaction and the $z$ axis corresponds now to the surface normal. Finally, note that the external magnetic field and the localized spin-orbit interaction that gives rise to the effective spin hamiltonian  are very weak compared to $\Delta v_s(\vec{r})\vec{\sigma}.\hat{n}$ in Eq.(\ref{eq:vdef}) and their effects on the variation of the  electronic states with the spin direction $\hat{n}$ has been neglected.

In many cases the variation of the electronic structure of the energy scale of the spin-wave excitations is negligible justifying an impulsive approximation for the IET. In analogy to the impulsive approximation for inelastic scattering~\cite{CheWic52}, the transition probability from the initial state of an incident electron in the tip with the spin of the adsorbate being in its  ground state $|\Phi_0\rangle$ to a final state of an outgoing electron in the sample with the spin of the adsorbate being in a state $|\Phi_n\rangle$ is obtained from matrix elements $T_{\mu\sigma^\prime n;\nu\sigma 0}$ of the tunneling amplitude $T_{\mu\sigma^\prime;\nu\sigma}(\hat{n})$ in Eq.(\ref{eq:TunAmp}) as,  
\begin{equation}
T_{\mu\sigma^\prime n;\nu\sigma 0}=\langle\Phi_n|T_{\mu\sigma^\prime;\nu\sigma}(\hat{n})|\Phi_0\rangle
\label{eq:MatDef}
\end{equation}
Taking into account the Pauli exclusion principle in the summation of tunneling probabilities over final electronic states, the resulting tunneling conductance will have a step $\Delta G_{n}$  at a bias $V$ corresponding to the spin-excitation energy $\Delta E_n = E_n-E_0$: $eV = \Delta E_n$. The magnitudes of these conductance steps are now obtained by replacing the matrix element in Eq. (\ref{eq:GDef}) by the matrix element in Eq. (\ref{eq:MatDef}) and are given by,
\begin{equation}
\Delta G_n = \frac{2\pi e^2}{\hbar} \sum_{\mu,\sigma^\prime; \nu,\sigma}
|T_{\mu\sigma^\prime n;\nu\sigma 0}|^2\delta(\epsilon_F -\epsilon_\mu)\delta(\epsilon_F -\epsilon_\nu)
\label{eq:DGRes}
\end{equation}
Note that the neglect of the small energy transfer and the use of the on-shell tunneling amplitude is consistent with the impulse approximation. Using the TH result for the tunneling amplitude in Eq.(\ref{eq:TunAmp}) and the matrix form of sample wave functions in Eq.(\ref{eq:BarPsiDef}), Eq.(\ref{eq:DGRes}) simplifies to
\begin{equation}
\Delta G_n = C_t \sum_{\mu} {\mathrm Tr}[
\langle\Phi_n|\bar{\psi}_{\mu} ( \vec{r}_0;\hat{n})|\Phi_0\rangle\langle\Phi_0|\bar{\psi}^\dagger_{\mu} ( \vec{r}_0;\hat{n})|\Phi_n\rangle] \delta(\epsilon_F - \epsilon_\mu)
\label{eq:GnExp}
\end{equation}
$\Delta G_n$ can now be evaluated using the explicit form for 
$\bar{\psi}_{\mu} ( \vec{r};\hat{n})$ in Eq.(\ref{eq:BarPsiRes}). The matrix elements in Eq.(\ref{eq:GnExp}) are then obtained as,
\begin{equation}
\langle\Phi_n|\bar{\psi}_{\mu} ( \vec{r}_0;\hat{n})|\Phi_0\rangle = 
\psi_{0\mu}(\vec{r}_0)\delta_{n0} +\psi_{s\mu}(\vec{r}_0)
\frac{\langle\Phi_n|\vec{S}|\Phi_0\rangle.\vec{\sigma}}{\sqrt{S(S+1)}}
\label{eq:MatRes}
\end{equation}
and from the elementary algebraic properties of the Pauli matrices, the trace of the matrix elements in Eq.(\ref{eq:GnExp}) simplifies to
\begin{equation}
{\mathrm Tr}[\langle\Phi_n|\bar{\psi}_{\mu} ( \vec{r}_0;\hat{n})|\Phi_0\rangle\langle\Phi_0|\bar{\psi}^\dagger_{\mu} ( \vec{r}_0;\hat{n})|\Phi_n\rangle] =
|\psi_{0\mu}(\vec{r}_0)|^2\delta_{n0} + |\psi_{s\mu}(\vec{r}_0)|^2\frac{| \langle\Phi_n|\vec{S}|\Phi_0\rangle|^2}{S(S+1)}
\label{eq:Tr}
\end{equation}
which gives the following simple expression for the inelastic steps in the conductance, 
\begin{equation}
\Delta G_{n} = C_t\left(\rho_0(\vec{r}_0,\epsilon_F)\delta_{n0} + 
\rho_s(\vec{r}_0,\epsilon_F)\frac{| \langle\Phi_n|\vec{S}|\Phi_0\rangle|^2}{S(S+1)}\right) .
\label{eq: DGnFin}
\end{equation}
Here $\rho_{0,s}(\vec{r},\epsilon_F)$ are LDOS, defined as,
\begin{equation}
\rho_{0,s}(\vec{r},\epsilon_F) = 2\sum_{\mu} |\psi_{0,s\mu} ( \vec{r})|^2\delta(\epsilon_ F- \epsilon_\mu) .
\end{equation}
The result for $\Delta G_{n}$ in Eq.(\ref{eq:GnExp}) is the main result of this paper and will now be discussed in some more detail.

At biases larger than the maximum spin-excitation energy, the conductance reduces to the conductance $G_0$ for a fixed spin. In this bias range, the conductance is given by,  
\begin{equation}
\sum_n \Delta G_n  = C_t\left(\rho_0(\vec{r}_0,\epsilon_F) + 
\rho_s(\vec{r}_0,\epsilon_F)\sum_n \frac{|\langle\Phi_n|\vec{S}|\Phi_0\rangle|^2 } 
{S(S+1)}\right) 
\label{eq:DGnGo}
\end{equation}
and using the identity,
\begin{equation}
\sum_n | \langle\Phi_n|\vec{S}|\Phi_0\rangle|^2 = S(S+1) ,
\label{eq:Ssum}
\end{equation}
which follows from the completeness relation, one 
obtains,
\begin{equation}
G_0 \equiv G(V=0) =  \sum_n \Delta G_n
\label{eq:G0Def}
\end{equation} 
Thus the step conductancies $\Delta G_n$ normalized to $G_0$ are given by,
\begin{equation}
\frac{\Delta G_n}{G_0}  = \gamma(\vec{r}_0) \frac{| \langle\Phi_n|\vec{S}|\Phi_0\rangle|^2}{S(S+1)} \ , n > 0
\label{eq:DGnGo}
\end{equation}
where the dimension-less strength parameter, 
\begin{equation}
\gamma(\vec{r}_0) = \frac{\rho_s(\vec{r}_0,\epsilon_F)}{\rho_0(\vec{r}_0,\epsilon_F) + \rho_s(\vec{r}_0,\epsilon_F)} ,
\label{eq:gamDef}
\end{equation}
only depends on the LDOS of the sample and is always less than unity. At low biases below the threshold for spin-excitations, the conductance $G_<$ is given by $\Delta G_{0}$ and 
\begin{equation}
\frac{G_<}{G_0} = 1 -\gamma(\vec{r}_0) (1 - \frac{| \langle\Phi_0|\vec{S}|\Phi_0 \rangle|^2}{S(S+1)}) .
\label{eq:G<Res}
\end{equation}
Thus our results for $\Delta G_{n}$ in Eqns (\ref{eq:DGnGo}) and (\ref{eq:G<Res}) justifies the analysis carried out by Hirjebedin et al. \cite{Hiretal07} of the observed step-like conductance by inelastic tunneling from the localized spin-excitations using $| \langle\Phi_n|\vec{S}|\Phi_0\rangle|^2$. In addition, our result gives a simple expression for the magnitude of $\frac{\Delta G_n}{G_0}$ and its lateral spatial variation with respect to the tip position, which can be calculated directly from spin-polarized wave functions at the Fermi level of the sample. 

The result for IET from a localized spin excitation has some important similarities with the result from IET from a localized vibration. The step in the conductance $\Delta G$ when exiting a local vibrational mode  of an adsorbate on a non-magnetic metal surface from its vibrational ground state to its first excited state was shown from perturbation theory and the TH approximation to be given by~\cite{LorPer00a,LorPer00b},
\begin{equation}
\Delta G = 2 C_t
\sum_{\mu}|\langle\vec{r}_0|\delta\psi_\mu\rangle|^2
\delta(\epsilon_F-\epsilon_\mu)
\label{eq:IETvib}
\end{equation}
where $|\delta\psi_\mu\rangle$ is the first order change in the spin-degenerate one-electron states, $|\psi_\mu(u)\rangle$, of the sample with respect to a root-mean-square amplitude  $\delta u = \sqrt{\hbar\over 2M\Omega}$ of the vibration with frequency $\Omega$, and mass $M$. This result for the conductance step can be written in a similar form to the corresponding result for the spin excitation in Eq.(\ref{eq: DGnFin}) but now using vibrational eigen-states $|\Phi_n\rangle$ as,
\begin{equation}
\Delta G = 2 C_t
\sum_{\mu}\langle\Phi_1|\langle\vec{r}_0|\psi_\mu(u)\rangle|\Phi_0\rangle\langle\Phi_0|\langle\psi_\mu(u)|\vec{r}_0\rangle|\Phi_1\rangle
\delta(\epsilon_F-\epsilon_\mu)
\end{equation}
Thus in this case the IET comes from a change of the amplitude of the tails of the wave functions in the vacuum region during the vibrational motion whereas in the case of a spin-excitation, the IET arises from the relative change in the amplitude between the two spin components along a fixed quantization direction of the vacuum tails of the spinor wave functions.   

Finally, we would like to make some comments on the underlying approximations and restrictions behind our result for the step-like conductance by inelastic tunneling from a localized spin excitation. The many electron effects on the IET are only introduced in the final state so that any influence of the opening of an inelastic channel on the elastic channel is neglected. In the case of IET from a vibrational excitation this influence can give rise to a decrease of the conductance instead of an increase~\cite{PerBar87}. Furthermore, the spin-dynamics is treated in the adiabatic approximation so that life time effects resulting in a broadening of the conductance steps is neglected. The results can be straightforwardly generalized to non-zero temperature and negative bias. In analogous manner for inelastic electron tunneling from vibrations, the Fermi surface smearing of the tip and the sample give rise to a temperature dependent broadening of the steps~with about $5k_BT$ \cite{LamJak68}. The results at negative bias is obtained by considering single-electron tunneling from the tip to the sample and gives that  tunneling conductance $G(-V) = G(V)$ is an even function with respect to the bias. The results can also be generalized to localized spin excitations of several adatoms and a non-magnetic tip but that and the application to specific systems are deferred to later work.

\begin{acknowledgments}
Support from the Swedish Science Research Council (VR) and
Institute for Surface and Interface Science (ISIS) at UC Irvine is
gratefully acknowledged. The author is indebted to discussions with D.L. Mills and Bechara Muniz. 
\end{acknowledgments}
 
\bibliography{MPref,MyPapers}

\end{document}